\documentclass[sigconf,nonacm]{acmart}

\usepackage{soul} 

\usepackage{subcaption}

\usepackage{listings, mdframed}
\usepackage{xcolor}  

\usepackage{color}
\usepackage[normalem]{ulem}
\usepackage{url}

\DeclareUrlCommand\ULurl{%
  \renewcommand\UrlLeft{\uline\bgroup}%
  \renewcommand\UrlRight{\egroup}}

\usepackage{enumitem} 

\usepackage{algpseudocode}
\usepackage{algorithm}
\usepackage{tabularx,multirow,booktabs,blindtext}
\usepackage{graphicx}
\usepackage{amsmath}
\usepackage{comment}
\usepackage{hhline}
\usepackage{bm} 

\lstdefinelanguage{YAML}{
  keywords={true,false,null,y,n, privileged},
  keywordstyle=\color{purple}\bfseries,
  basicstyle=\footnotesize\ttfamily,
  morecomment=[l]{\#},
  commentstyle=\color{purple}\ttfamily,
  stringstyle=\color{red}\ttfamily,
  morestring=[b]',
  morestring=[b]"
}

\algnewcommand\algorithmicforeach{\textbf{for each}}
\algdef{S}[FOR]{ForEach}[1]{\algorithmicforeach\ #1\ \algorithmicdo}

\newcommand{\name}{GenKubeSec\xspace}

\newcommand{\nameFirstComponent}{GenKubeDetect\xspace}

\newcommand{\nameSecondComponent}{GenKubeResolve\xspace}

\AtBeginDocument{%
  }

\acmConference[ACSAC '24]{Make sure to enter the correct conference title from your rights confirmation emai}
{December 09--13, 2024}{Waikiki, Hawaii, USA}
\acmISBN{978-1-4503-XXXX-X/24/12}

\begin{document}

\title{\name: LLM-Based Kubernetes Misconfiguration Detection, Localization, Reasoning, and Remediation}

\author{Ehud Malul, Yair Meidan, Dudu Mimran, Yuval Elovici, Asaf Shabtai}
\affiliation{%
  \institution{Department of Software and Information Systems Engineering \\ Ben-Gurion University of The Negev}
  \country{}
}

\renewcommand{\shortauthors}{Malul et al.}

\begin{abstract}
  A key challenge associated with Kubernetes configuration files (KCFs) is that they are often highly complex and error-prone, leading to security vulnerabilities and operational setbacks.
  Rule-based (RB) tools for KCF misconfiguration detection rely on 
  static rule sets, making them inherently limited and unable to detect newly-discovered misconfigurations.
  RB tools also suffer from misdetection, since mistakes are likely when coding the 
  detection rules. 
  Recent methods for detecting and remediating KCF misconfigurations are limited in terms of their scalability and detection coverage, or due to the fact that they have high expertise requirements and do not offer automated remediation along with misconfiguration detection.  
  Novel approaches that employ LLMs in their pipeline rely on API-based, general-purpose, and mainly commercial models. 
  Thus, they pose security challenges, have inconsistent classification performance, and can be costly. 
  In this paper, we propose \name, a comprehensive and adaptive, LLM-based method, which, in addition to detecting a wide variety of KCF misconfigurations, also identifies the exact location of the misconfigurations and provides detailed reasoning about them, along with suggested remediation.  
  When empirically compared with three industry-standard RB tools, \name achieved equivalent precision ($0.990\pm0.020$) and superior recall ($0.999\pm0.026$). 
  When a random sample of KCFs was examined by a Kubernetes security expert, \name's explanations as to misconfiguration localization, reasoning and remediation were 100\% correct, informative and useful.
  To facilitate further advancements in this domain, we share the unique dataset we collected, a unified misconfiguration index we developed for label standardization, our experimentation code, and \name itself as an open-source tool. 
  A video demonstrating our implementation of \name can be found here: \ULurl{https://youtu.be/hBehYfdR-zM}.
\end{abstract}



\received{28 May 2024}
\received[revised]{7 August 2024}
\received[accepted]{19 August 2024}

\maketitle

\section{Introduction}\label{sec:Introduction}

The cloud native computing landscape has undergone a significant transformation, with rapid adoption of container-based environments in a relatively short period of time. 
Kubernetes (K8s)~\cite{kubernetes2019kubernetes} has emerged as the preferred choice for orchestrating containerized applications, driven by the benefits K8s offers, such as efficiency, portability, isolation, scalability, and flexibility~\cite{CTOclaim}. 
The strong community~\cite{k8sCommunity} support surrounding containerization technologies has strengthened their leading role in the development ecosystem.

The plethora of K8s configuration files (KCFs) available in public repositories~\cite{artifacthub} streamlines the creation and deployment of new environments and applications, permitting developers to easily create cloud environments by duplicating open-source KCFs. 
However, this convenience comes with its own set of challenges.
Open-source KCFs, while widely available, may contain undetected bugs, errors, vulnerabilities, and other types of misconfigurations (abbreviated as misconfigs), potentially leaving deployed systems exposed to cyber threats~\cite{rangta2022tools,cisBenchmark,tripathi2024attacking,haimed2023exploiting,koushki2024root,guffey2023cloud}. 
Moreover, due to the adoption of K8s in a broad variety of production environments, and the accessibility of open-source KCFs, attackers invest significant efforts in identifying KCF misconfigs that can serve as exploitation entry points into clusters or applications~\cite{konala2023sok,microsoftBlog}.

Given the associated risks, powerful tools capable of detecting KCF misconfigs are required. 
Rule-based (RB) static analysis tools like SLI-Kube~\cite{rahman2023security}, Checkov~\cite{checkov}, KubeLinter~\cite{kubelinter}, and Terrascan~\cite{terrascan} are recognized tools for assessing the security of KCFs based on predefined rules.
However, due to their static nature, 
RB tools also struggle with adapting to new misconfigs, primarily due to the need for coding new error-prone detection rules. 
Recent studies on KCF misconfig detection proposed methods that rely on more innovative approaches, such as knowledge graphs and topological graphs~\cite{haque2022kgsecconfig,blaise2022stay}.
However, these methods face limitations in scalability, expertise requirements, reliance on high-quality input data, and automated remediation, as elaborated in Secs.~\ref{subsec:KCF_misconfig_detection} and~\ref{subsec:KCF_misconfig_remediation}.

Recent advances in large language models (LLMs) have proven highly effective in various application domains~\cite{goyal2024healai,hu2024bliva}, including cybersecurity~\cite{sewak2023crush,yao2024survey}, where several studies suggested using LLMs in static analysis tasks~\cite{kwon2023exploring,yu2024security,ahmad2024hardware,liu2023harnessing,li2023assisting}.
Since KCFs are technically (semi-structured) text files, two studies focused specifically on LLM-based methods for KCF misconfig detection and remediation~\cite{minna2024analyzing,lanciano2023analyzing}.
These studies, however, have the following limitations: (1) each of them utilized LLMs to solve only a sub-task rather than addressing all of the related challenges from end-to-end, including detection, localization, reasoning, and remediation, (2) they utilized pretrained LLMs which had only undergone basic domain adaptation and not LLMs fine-tuned to the specific task, thus demonstrating limited performance, (3) the inspected files were sent to external LLMs via an API, introducing security (privacy) risks and adding unnecessary overhead, (4) relatively small datasets with small number of misconfig types were used for evaluation, and (5) the experimental results were reported very scarcely, so their performance during future inference cannot be guaranteed.

In light of the need to mitigate KCF-related risks, the limitations of existing RB and graph-based methods, the potential of LLMs, and the shortcomings and scarcity of research on applying LLMs in KCF misconfig detection and remediation, in this paper we introduce \emph{\name}: a novel and comprehensive LLM-based method which comprises of three key components. 
In the first component, a substantial set of KCFs is collected, and then labeled and standardized using a unified misconfig index we developed. 
In the second component, named \emph{\nameFirstComponent}, since general purpose pretrained LLMs performed poorly, a base LLM is optimized and fine-tuned using a large and comprehensive set of labeled KCFs to detect a wide variety of KCF misconfigs (the scope of which continually grows over the course of time). 
In the third component, named \emph{\nameSecondComponent}, a pretrained LLM is adapted, using prompt engineering and few-shot learning techniques~\cite{brown2020language,lin2022few,zhang2023machine}, to provide the exact location of each misconfig detected in a KCF by \nameFirstComponent, clear explanations in human language, and suggested remediation.
Unlike previous studies where LLMs were utilized for either KCF misconfig detection~\cite{lanciano2023analyzing} or remediation~\cite{minna2024analyzing}, we propose an end-to-end LLM-based method which addresses both of these aspects. 
In addition, \nameSecondComponent also provides localization and reasoning for each misconfig detected by \nameFirstComponent. 

In our extensive evaluation, \nameFirstComponent achieved detection precision of 0.990 (equivalent to three industry standard RB tools) and recall of 0.999 (superior to any of those tools individually). 
In addition, manual, expert-based analysis of false misconfig detections made by \nameFirstComponent revealed that more than 83\% of these detections were actually correct.
This demonstrates the ability of our LLM-based method to generalize and detect 
misconfigs which are not covered by the rule sets of the three RB tools. 
These evaluation results were achieved with a relatively lightweight LLM architecture and thus save substantial time and compute.
Another benefit is that \nameFirstComponent supports KCF misconfig learning both in primary batch mode, as part of fine-tuning, and in ongoing mode, using just a few examples of newly-discovered misconfigs.
With \nameSecondComponent, based on a random sample of KCFs, a K8s expert validated that \emph{all} of the localization, reasoning, and remediation suggestions were clear and accurate.
A notable advantage of \name, is that by relying on local LLMs rather than API-based LLMs, \name minimizes security and privacy risks, as it does not expose proprietary KCFs (with potential vulnerabilities) to external entities.  
To conclude, the contributions of our research are as follows:

\begin{itemize}[nosep, leftmargin=*]
    \item To the best of our knowledge, we are the first to propose an end-to-end LLM-based method that provides KCF misconfig detection (both known and new), along with localization, reasoning, and remediation. 
    \item As part of our quantitative evaluation, we developed a unified misconfig index (UMI) of 169 KCF misconfigs to enable standardization among detection tools; this index can be used by the scientific community and industry to further advance research in this area through improved consistency and comparability.
    \item We collected a large 
    and diverse set of real-world material from the K8s domain. 
    This dataset includes approximately 700MB of K8s-focused free text, as well as almost 277,000 KCFs which we labeled using the abovementioned UMI. 
    Such datasets are known to be rare~\cite{lanciano2023analyzing}, and therefore, we make this unique dataset publicly available to enable our research to be reproduced, and to advance research in this area by the scientific community.
    \item To further facilitate transparency, usability, and benchmarking, we also make our code publicly available via a GitHub repository and make \name available as an open-source tool.
\end{itemize}

\begin{figure}[t]
    \begin{lstlisting}[language=YAML]
    apiVersion: v1
    kind: Pod
    metadata:
      name: pod-name
    spec:
      containers:
        - name: some-container
          image: some-image
          command: [ some-command ]
      securityContext:
        privileged: true
    \end{lstlisting}
    \vspace{-0.4cm}
    \caption{An example of a misconfigured KCF: the container is configured to run with privileged access; this can potentially expose sensitive host information. 
    }
    \label{fig:KCF_misconfig_example}
    \Description{An example of a KCF with a critical security flaw - a container that runs with privileged access, potentially exposing sensitive host information.}
\end{figure}

\section{Background}\label{sec:Background}

\subsection{K8s and related security challenges}\label{subsec:K8s_and_related_security_challenges}

\subsubsection{K8s and KCFs}\label{subsubsec:KCFs}
K8s is the primary container orchestration tool for 71\% of Fortune 100 companies~\cite{CNCFreport,k8sCreation}. 
KCF (also known as a \emph{manifest file}~\cite{kubernetesManifest}), refers to any file that defines the configuration of various K8s components~\cite{MediumArticleOnManifest}, and describes how an application or service should be deployed and run within a K8s cluster~\cite{KubernetesManifestDocs}. 
KCFs provide a 
human-readable representation (in a YAML or JSON format) of an application's requirements, enabling reproducible deployment and consistent application behavior. 
KCFs are essential building blocks in containerized and cloud-native architectures, enabling developers and operators to define, manage, and scale applications effectively. 

\begin{figure*}[ht]
    \centering
    \includegraphics[width=0.95\linewidth]{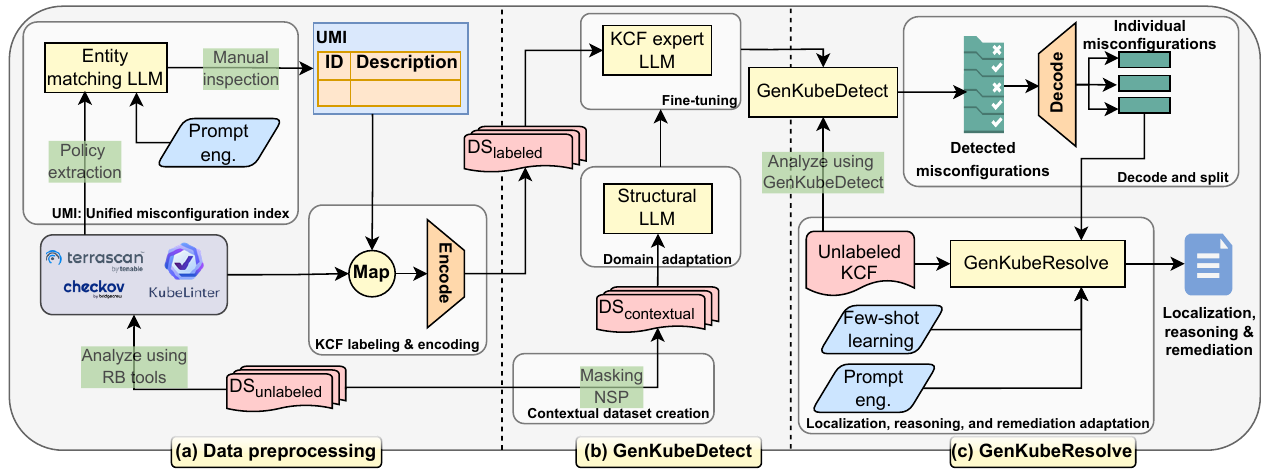}
    \caption{Overview of \name. 
    This method includes a data preprocessing component (a), in which a unified misconfig index (UMI) is created, and KCFs are labeled and encoded; the \nameFirstComponent component (b) in which an LLM for KCF structure understanding and misconfig detection is trained;  the \nameSecondComponent component (c) in which a pretrained LLM with prompt engineering and few-shot learning is used for localization, reasoning, and remediation of identified misconfigs.}
    \label{fig:method}
\end{figure*}

\subsubsection{KCF misconfigs and vulnerabilities}\label{subsubsec:KCF_misconfigs_and_vulnerabilities}

The complexity associated with configuring KCFs mainly stems from the inter-dependencies among the large variety of configuration parameters~\cite{samir2023adaptive}. 
This complexity can lead to misconfigs in KCFs. 
A KCF misconfig (see example in Fig.~\ref{fig:KCF_misconfig_example}) is the result of the incorrect configuration of system components.
Such misconfig can lead to security vulnerabilities and affect system workload, integrity, and performance. 
KCF misconfigs can occur at the edge, application, or cluster level, allowing attackers to exploit containers and hosts.
They can also occur with parameters affecting resource allocation, networking, security policies, and more. 
These misconfigs can have far-reaching consequences, resulting in application downtime, performance bottlenecks, and resource wastage.
KCF misconfigs may also lead to non-compliance with industry regulations and data protection laws~\cite{KubernetesMisconfigurationPost}.
According to a CNCF report from 2022~\cite{CNCFNumberOneRiskIsMisconfiguration}, KCF misconfigs are the biggest security threat when using K8s. 
This report also states that 94\% of companies that use K8s encountered a security vulnerability. 
Another study~\cite{ARMO} revealed that 100\% of 10,000 checked clusters had misconfigs, and 65\% had at least one high-severity misconfig.

\subsection{Large language models}\label{subsec:Large_language_models}

LLMs represent a significant advancement in natural language processing (NLP), enhancing machines' ability to understand and generate human-like language. 
Utilizing deep learning (DL) techniques and extensive textual datasets, LLMs excel in tasks like sentiment analysis~\cite{li2019exploiting} and text generation and translation~\cite{zhang2023survey}. 
Contemporary pretrained LLMs, which continuously evolve and increase in size and sophistication, can be adapted for specialized domains~\cite{hadi2023large} using, e.g., fine-tuning and continual learning, as described next.

Fine-tuning a pretrained LLM involves additional training focused on specific objectives, like aligning model outputs with user intent, generating ethical and accurate responses, and improving performance on particular tasks. 
This process not only enhances the model's ability to understand and respond to task-specific instructions but also makes it more robust to domain shifts, without significantly increasing computational costs. 
Fine-tuning methods vary and include transfer learning with task-specific data, instruction-tuning using formatted data, and alignment-tuning to align model outputs with human values.
These approaches ensure that LLMs operate effectively in varied and specific contexts~\cite{naveed2023comprehensive}.

Continual learning (CL)~\cite{wu2021pretrained} is a pivotal aspect of modern machine learning (ML). 
CL enables systems to gradually learn and integrate diverse contents or skills via an additional step of training, using newly-acquired data.  
This approach is essential for developing robust and adaptable ML systems capable of responding to changing demands and incorporating new knowledge efficiently~\cite{wang2023comprehensive}.
In the context of domain adaptation, CL enhances an LLM's ability to balance the retention of original domain knowledge with learning from the targeted domain distribution~\cite{alyafeai2020survey,naveed2023comprehensive, sun2019fine}, ensuring effective adaptation without `catastrophic forgetting'~\cite{french1999catastrophic,luo2023empirical}.
Moreover, CL maintains computational efficiency during domain-adaption~\cite{rongali2020continual}.
Employing techniques like next sentence prediction (NSP) which help the model maintain context and coherence between sequences, and masking, which focuses the model's attention on relevant parts of the input~\cite{araujo2023learning,devlin2018bert}, along with unsupervised learning~\cite{tyagi2022unsupervised, wilson2020survey}, enhance LLMs' comprehension and response capabilities.

\section{Proposed method}\label{sec:proposed_method}

In this paper, we propose \name: a novel LLM-based method for KCF misconfig detection, localization, reasoning, and remediation, aimed at mitigating KCF-related K8s security vulnerabilities. 
Our method consists of three components. 
Initial data preprocessing is performed in the first component, and LLMs are used in the other two components as follows: 
The \nameFirstComponent component is trained (fine-tuned) to detect any misconfigs that may exist in a given KCF, serving as an advanced, LLM-based alternative to existing rule- and graph-based tools and methods (described in Sec.~\ref{subsec:KCF_misconfig_detection}). 
The \nameSecondComponent component, which is based on a pretrained LLM and specifically crafted prompts, is trained to provide localization, reasoning, and remediation suggestions for each misconfig detected by \nameFirstComponent.

\subsection{Data preprocessing component}\label{subsec:Data_preprocessing}

\subsubsection{Creating a unified misconfig index}\label{subsubsec:creating_a_unified_misconfig_index}

Given the large number of KCF misconfig types that can be detected by various tools, recent studies~\cite{minna2024analyzing,haq2024lucid} stressed the need for misconfig label standardization. 
To address this need, we createded a unified misconfig index (UMI).
Our UMI consists of multiple, consistently-formatted, unique misconfig class labels. That is, if a certain misconfig was annotated differently in various sources, whether described differently in text or using different misconfig IDs, the UMI maps all of the variations of the misconfig to a single ID (i.e., label). In addition to being useful in our research, the UMI can facilitate the development of effective solutions and their comparison.

The first step in creating the UMI involved collecting as many policies and detection rules as possible from multiple reliable resources, which in our case were the three RB KCF misconfig detection tools: Checkov, KubeLinter, and Terrascan. 
In the second step, an LLM was employed for entity matching~\cite{peeters2023entity,peeters2023using} using prompt engineering (see Fig.~\ref{fig:method}.a); the entity matching prompt structure is presented in Appendix~\ref{app:Entity_matching_prompt_structure}. 
Manual inspection of the entity matching results was then performed, to verify their correctness; this was followed by manual adjustments when needed. 
Entries in the UMI (each of which represents a KCF misconfig) are formatted as the combination of a numerical <misconfig\_ID> $\in (0, 1, 2, \dots)$ and a textual <misconfig\_description>, for example, (9, 'Default namespace should not be used`). 
The UMI we developed can be found in our GitHub repository.

\subsubsection{Collecting and labeling KCFs}\label{subsubsec:Collecting_KCFs_and_labeling_them} 
To train the LLMs used in \nameFirstComponent and \nameSecondComponent, we collect a large and diverse set of KCFs. 
We annotate this raw dataset of unlabeled KCFs as $DS_{unlabeled}$ (see Fig.~\ref{fig:method}.a). 
Then, to create $DS_{labeled}$, we associate ground-truth misconfig labels to the KCFs in $DS_{unlabeled}$ using Checkov, KubeLinter, and Terrascan (which the UMI is based on). 
We opted to use three detection tools due to the limitations of RB detection methods.
Nonetheless, $DS_{labeled}$ may still consist of some undetected misconfigs that are not covered by the unified rule sets of the three RB tools.
In practice, when any of these tools detects a misconfig for a KCF in $DS_{unlabeled}$, the detected misconfig is mapped to its corresponding UMI misconfig\_ID
and replaced by a pair of values formatted as (<impacted\_K8s\_resource>+<misconfig\_ID>), e.g., `app+52' (see full example in Appendix~\ref{app:Example_of_Encoded_and_decoded_labels}).
We consider each pair of such values as an encoded label. 
Encoding (compressing) the class labels decreases the overall size of labeled KCFs, such that (1) higher coverage is provided by LLMs that impose size limits
, (2) training is accelerated, and (3) fewer misclassifications are likely to occur, as shorter class labels are easier for the LLM to (re-)generate.

\subsection{The \nameFirstComponent component}\label{subsec:Component_1_Misconfig_detection}

The \nameFirstComponent component of \name, is a fine-tuned LLM which specializes in detecting KCF misconfigs. 
Specifically, given a KCF as input, \nameFirstComponent's main goal is to use an LLM in order to generate a list of detected misconfigs, formatted as the encoded labels described above. 
We achieve this goal by selecting a base LLM (which is preferably pretrained on English semi-structured data to resemble the hierarchical textual format used by KCFs) and fine-tuning it for the task of misconfig detection. 
Fine-tuning the LLM is performed in two stages: (1) adapting it so that it `understands' the general structure of KCFs, and (2) further adapting it so that it can detect misconfigs in KCFs.

\subsubsection{Adapting the base LLM for KCF structure understanding}\label{subsubsec:Fine_tuning_the_base_LLM_for_KCF_structure_understanding}

To familiarize \nameFirstComponent's LLM with the specific structure of KCFs, it is fine-tuned using the $DS_{contextual}$ dataset (Fig.~\ref{fig:method}.b). 
$DS_{contextual}$ is constructed by applying two techniques on $DS_{unlabeled}$, namely masking and NSP (see Sec.~\ref{subsec:Large_language_models}).
Masking is randomly applied to 15\% of each KCF (we chose to mask 15\%, as this is typically done in similar cases~\cite{devlin2018bert, raffel2020exploring, liu2019roberta}). 
Using NSP, \nameFirstComponent is trained to predict the second half of a KCF based on its first half.
This first stage of the fine-tuning process enables the LLM to recognize and interpret the details and dependencies inherent in KCFs, thereby enhancing its ability to perform subsequent domain-specific tasks (such as misconfig detection) with greater accuracy and efficiency.
The output of this stage is the structural LLM (shown in Fig.~\ref{fig:method}.b).

\subsubsection{Fine-tuning the structural LLM for KCF misconfig detection}\label{subsubsec:Fine_tuning_the_base_LLM_for_KCF_misconfig_detection}
Once specialized on the structure of KCFs, the structural LLM is further fine-tuned using $DS_{labeled}$
, enabling it to function as a multi-label classifier; thus, given a KCF as input during inference, the LLM will generate text that serves as the predicted class labels of (zero or more) detected misconfigs.
To enhance this process, we used Low-Rank Adaptation (LoRA)~\cite{hu2021lora, huggingfaceLora}. LoRA is a technique that allows fine-tuning a subset of parameters within a large pre-trained model by decomposing a large matrix into two smaller low-rank matrices in the attention layers, while keeping the original weights unchanged. This makes the process more efficient, as it requires fewer resources and allows for faster convergence without the need to retrain the entire model from scratch. LoRA preserves the integrity of the original model's weights, maintaining its knowledge and capabilities while adapting to specific tasks or datasets.
The output of this stage is the KCF expert LLM (shown in Fig.~\ref{fig:method}.b).

\subsection{The \nameSecondComponent component}\label{subsec:Component_2_Misconfig_localization_reasoning_remediation} 

Given a KCF as input, 
\nameFirstComponent is trained to generate (possibly multiple) detected misconfig labels, each in the previously mentioned encoded format, e.g., `app+52.'
Then, to expand GenKubeSec's capabilities further, for each misconfig detected by \nameFirstComponent, \nameSecondComponent is trained to generate (1) the exact location of the misconfig, (2) reasoning regarding the nature or cause of the detected misconfig, and (3) recommended remediation steps.
To accomplish this, in the first stage of this component we decode and split the class labels, and in the second stage we adapt a pretrained LLM using prompt engineering techniques, as described below.

\subsubsection{Decoding and splitting}\label{subsubsec:decoding_splitting} 

In the first stage of \nameSecondComponent, the encoded labels are decoded (see Fig.~\ref{fig:method}.c) and then split into pairs of (inspected\_KCF, decoded\_label). 
In practice, we replace the misconfig\_ID in each encoded label with the associated misconfig\_description from the UMI (see Appendix~\ref{app:Example_of_Encoded_and_decoded_labels}); this provides the LLM with much more contextual information, enabling it to respond more accurately.
The subsequent splitting (illustrated in Fig.~\ref{fig:method}.c) performed at this point is aimed at preventing the LLM from being `confused' due to an overabundance of misconfigs in the same KCF. Moreover, label splitting enables this component to provide localization, reasoning, and remediation suggestions for each misconfig separately.
    
\subsubsection{Adapting an LLM for localization, reasoning, and remediation}\label{subsubsec:Specializing_a_LLM_for_Localization_reasoning_and_remediation}

The task of detecting a large variety of K8s misconfigs performed by \nameFirstComponent is an extremely specific task for a general-purpose pretrained LLM, (as reflected by the experimental results in Sec.~\ref{subsec:Evaluating_Existig_pre_trained_LLMs}).
This is why we chose to implement \nameFirstComponent using fine-tuning techniques; these techniques, however, impose significant overhead.
In contrast, preliminary experimentation with \nameSecondComponent showed that for the task of providing localization, reasoning, and remediation for detected KCF misconfigs, fine-tuning is not essential.
Instead, we decided to adapt a pretrained LLM for this task, using few-shot learning techniques (which provide the LLM with favorable relevant examples as part of in-context learning~\cite{brown2020language,lin2022few,zhang2023machine}) and prompt engineering techniques. 
An illustration of this process can be found in Fig.~\ref{fig:method}.c, while prompt examples are provided in Appendix~\ref{app:nameSecondComponent_Model_Inputs_and_Outputs_for_Misconfig_Remediation_Localization_and_Explanation}.

The base model we chose for \nameSecondComponent is a pretrained Mistral LLM. 
It was chosen due to its superior recall and F1-score (abbreviated as F1) in our preliminary experiments (Sec.~\ref{subsec:Evaluating_Existig_pre_trained_LLMs}), the fact that it is one of the best open-source LLMs available~\cite{jiang2023mistral}, and the ability to use it locally, free of charge. 
Local use of Mistral ensures that, unlike web-based pretrained LLMs, no KCF data is exposed to external APIs. 
Using a local LLM also accelerates operation by eliminating the delays and complexities of API calls.

To specialize Mistral for localization, reasoning, and remediation, we optimized a system prompt which, based on a provided KCF and a misconfig detected in it, instructs Mistral to return the exact line number of the misconfig, the logic behind the detection, and suggestions on how to fix the misconfig. 
To enhance Mistral's performance, we augmented this system prompt with a few examples of KCFs, detected misconfigs, and expected output (i.e., few-shot learning). 
The system prompt we optimized and an inference example for \nameSecondComponent are presented in Appendix~\ref{app:nameSecondComponent_Model_Inputs_and_Outputs_for_Misconfig_Remediation_Localization_and_Explanation}.

\section{Evaluation method}\label{sec:evaluation_method}

\subsection{Experimental setup}\label{subsec:Experimental_setup}

Throughout our quantitative evaluation we used the Python programming language. 
The Hugging Face transformers library~\cite{wolf2020transformers} served as the primary source for training and utilizing state-of-the-art LLMs (see the complete code in the project's GitHub repository). 
For heavy-lifting computing, mainly during the LLM fine-tuning step, we used NVIDIA RTX 4090~\cite{rtx4090} and RTX 6000~\cite{rtx6000} Ada GPUs. 

To create our UMI (Sec.~\ref{subsubsec:creating_a_unified_misconfig_index}), we began by performing targeted web crawling of the online documentation (i.e policy indexes) of three open-source RB tools, namely KubeLinter, Checkov, and Terrascan, which are industry standards~\cite{k8sbesttools,k8sbesttools2}. 
We then utilized OpenAI's GPT-4 for entity matching, and establishing in a UMI of 170 standardized unique misconfigs, one of which was used for KCFs where no misconfig was identified. 

To produce $DS_{unlabeled}$, we collected 276,520 unlabeled KCFs from the K8s Manifest Subset~\cite{k8sYamlFiles}, recently released by `The Stack'~\cite{Kocetkov2022TheStack}. 
In order to associate ground-truth labels to each KCF in $DS_{unlabeled}$, we utilized the three open-source RB tools mentioned earlier (KubeLinter, Checkov, and Terrascan), resulting in $DS_{labeled}$.

\subsection{Data partitioning and performance metrics}\label{subsec:Data_partitioning_and_performance_metrics} 

To 
compare the performance of various KCF misconfig detectors, whether rule- or LLM-based, we randomly split $DS_{labeled}$ into fixed training (80\%), validation (10\%), and test (10\%) sets. 
We compared their performance using the common classification metrics of precision, recall, and the F1, weighted by the occurrence of the various KCF misconfigs in the test set. 
For convenience, in the remainder of the paper the terms precision, recall, and F1 represent the weighted precision, weighted recall, and weighted F1, respectively. 
To compute the value of these classification performance metrics, we define their building blocks as follows:

\textbf{True positives (TPs):} Instances where both the LLM and at least one RB tool correctly detected a misconfig.

\textbf{False positives (FPs):} Instances where the LLM erroneously detected a misconfig that was not recognized by any of the RB tools.
Note that due to the fact that RB detectors are limited by their given set of rules, it might be the case that a FP is actually a detection of a \emph{variant} of a known misconfig, such that it does not entirely match any of the coded detection rules. 

\textbf{False negatives (FNs):} 
Instances where a misconfig was overlooked by the LLM but detected by one of the RB tools.

\textbf{True negatives (TNs):} Instances in which there was consensus between the LLM and the RB tools as to the absence of misconfig.

\section{Experimental results}\label{sec:Experimental_results}

\subsection{Investigating pretrained LLMs}\label{subsec:Evaluating_Existig_pre_trained_LLMs}

The first experiment aimed to answer the following question: Are existing pretrained LLMs sufficient for detecting misconfigs in KCFs? 
To answer this question we experimented with four state-of-the-art pretrained LLMs: GPT-4-Turbo~\cite{achiam2023gpt, ChatGPT}, Gemini 1.0 Pro~\cite{pichai2023introducing, GeminiPro}, Claude 3 Sonnet~\cite{claude, ClaudeSoonet}, and Mistral-7B-Instruct-v0.2~\cite{jiang2023mistral, MistralModel}. 
Currently these models can be used via online chat or API calls but cannot be fine-tuned, so we relied on prompt engineering and few-shot learning~\cite{brown2020language,lin2022few,zhang2023machine}.
First, we created a fixed few-shot training set by randomly sampling 10 labeled KCFs (the number of labeled KCFs is based on Gemini's maximum allowed input size (in tokens)). 
To increase variety, we ensured that each of the 10 KCFs had at least two misconfig labels. 
The same few-shot training set was used with each pretrained LLM. 
Similarly, the same test set (consisting of 30 randomly sampled test KCFs with the same 39 misconfigs found in the training set) was used to evaluate the LLMs' performance. 
The prompt we used in this evaluation is presented in Appendix~\ref{app:Few_shot_learning_example}). 

\begin{figure}[ht]
    \centering
    \includegraphics[width=0.85\linewidth]{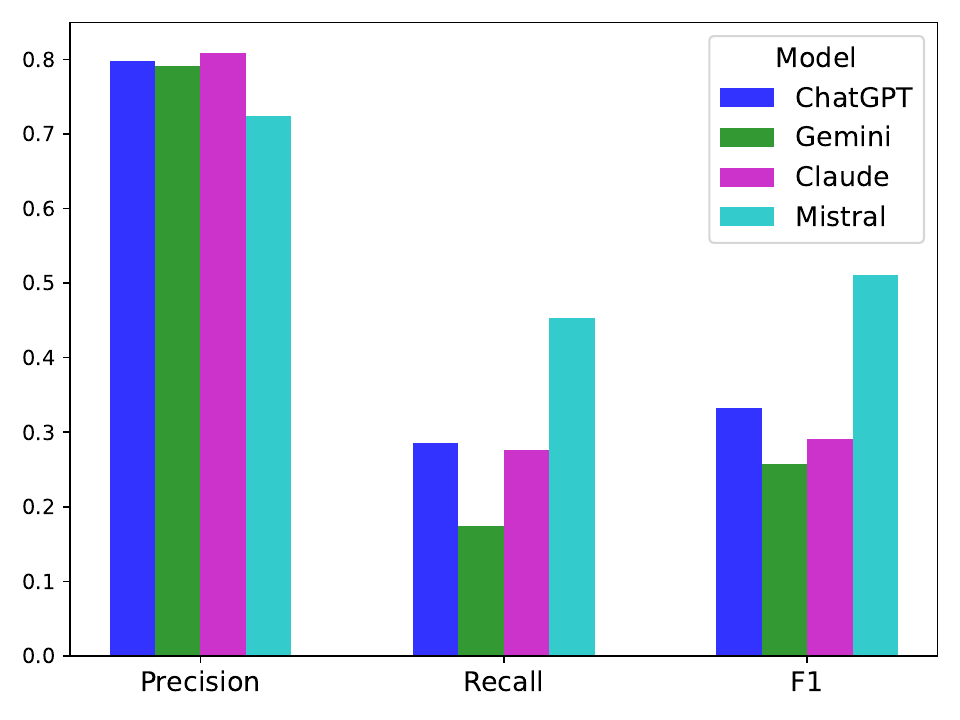}
    \caption{Performance of the pretrained LLMs using few-shot learning for misconfig detection.}
    \label{fig:results_existing_pre_trained_LLMs}
    \Description{Bar chart comparing the performance of four different pretrained LLMs.}
\end{figure}

As can be seen in Fig.~\ref{fig:results_existing_pre_trained_LLMs}, our experiment with the pretrained LLMs yielded unsatisfactory results: the highest precision, recall, and F1 values obtained were around 0.81, 0.45, and 0.51, respectively. 
Based on this, we conclude that the task of detecting misconfigs in KCFs might be too narrow for a pretrained general-purpose LLM (as concluded in past studies~\cite{chen2023can, cheshkov2023evaluation}) and the ability to realize LLMs' potential in this area might require more than the use of a limited set of labeled KCFs in a few-shot learning scheme. 
In Fig.~\ref{fig:results_existing_pre_trained_LLMs} we can also see that the recall obtained with Mistral is much higher than that achieved by the other pretrained LLMs, while its precision is lower. 
In classical ML settings, a combination of low precision and high recall typically indicates a classification threshold that is too low~\cite{davis2006relationship}, i.e., the ML model predicts too many positive labels. 
However, in our LLM settings, we found that Mistral tends to generate responses that are 2-3 times longer than those of the other LLMs, possibly due to the internal configuration of hyperparameters. 
In our use case, longer responses necessarily means higher positive misconfig detections (and descriptions), some of which are TPs while others are FPs, resulting in the low precision and high recall obtained by Mistral.

\subsection{Selecting the modeling architecture and base model}\label{subsec:Selecting_the_modeling_architecture_and_base_model}

Given the results obtained in the previous experiment and the need to achieve better performance and broader misconfig coverage, in this subsection, we explore the use of dedicated open-source LLMs with different architectures. 
These LLMs are optimized and fine-tuned for our task of misconfig detection in KCFs.

We considered three popular LLM architectures: the encoder-only, decoder-only, and encoder-decoder~\cite{devlin2018bert,radfordlanguage,vaswani2017attention}. 
Among them, we expected the encoder-decoder architecture to be the most suitable for our use case, because (1) the encoder-only architecture excels in context understanding~\cite{huggingfaceEncoder}, (2) the decoder-only architecture is well suited for generating tokens~\cite{huggingfaceDecoder}, and (3) the encoder-decoder architecture performs well at both understanding the context and generating appropriate tokens~\cite{huggingfaceEncoderDecoder}. 
Recently, the superiority of the encoder-decoder architecture in a code understanding and completion task was demonstrated~\cite{wang-etal-2021-codet5}; our use case is similar in that (1) a KCF has to be well understood by the LLM, and (2) LLM completion is then used to generate appropriate misconfig labels. 

To identify the most promising LLM architecture and base model for KCF misconfig detection, we experimented with several open-source LLMs with the abovementioned architectures (i.e., encoder-only, decoder-only, and encoder-decoder). 
For efficiency, we fine-tuned each of the LLMs using a stratified random sample of 68,997 labeled KCFs, which represent 30\% of the total 229,989 labeled KCFs that comply with the LLMs' token limits. 
Of these, 85\% were allocated for training and 15\% for validation. To enable comparison across our experiments, from this point onwards we utilized the same test set defined in Sec.~\ref{subsec:Data_partitioning_and_performance_metrics} (which comprises 22,987 instances that comply with the token limits). 
As can be seen in Table~\ref{tab:architecture_selection}, BigBird (encoder-only) and CodeLlama (decoder-only) failed to converge after fine-tuning, i.e., their output (generated text that serves as KCF labels) did not match the desired misconfig labels, thus we were unable to calculate the metrics for them.
In contrast, the other LLMs (which were mainly encoder-decoders, as expected) demonstrated adaptability to the designated task: their output gradually resembled the misconfig labels from the training set, and they managed to obtain noteworthy precision.

\begin{table}[t]
    
    \resizebox{\columnwidth}{!}{
    \centering
    \begin{tabular}{lllll}
        \toprule
        Base model & Architecture &Precision & Recall & F1\\
        \midrule
        BigBird~\cite{zaheer2020bigbird} & Encoder-only & --- & --- & ---\\
        CodeLlama~\cite{roziere2023code} & Decoder-only & --- & --- & --- \\
        CodeGen~\cite{nijkamp2022codegen} & Decoder-only & $0.982 \pm $0.040	 & $0.574 \pm $ 0.199	 & $0.705 \pm $0.168\\
        LED~\cite{Beltagy2020Longformer} & Encoder-decoder & $0.976 \pm $0.035	 & $0.265 \pm $0.248 & $0.368 \pm 0.256$\\
        CodeT5p 770M~\cite{wang2023codet5+} & Encoder-decoder & $\textbf{0.987} \pm $\textbf{0.023} & $\textbf{0.997} \pm $\textbf{0.018}	 & $\textbf{0.992} \pm $\textbf{0.018}\\
        CodeT5p 6B~\cite{wang2023codet5+} & Encoder-decoder & $0.902 \pm $0.068 & $0.390 \pm $0.015	 & $0.545 \pm $0.026 \\
        \bottomrule
    \end{tabular}
    }
    \caption{Performance with various LLM architectures}
    \label{tab:architecture_selection}
\end{table}

Of the evaluated encoder-decoder base models, we chose to proceed with CodeT5p 770M (trained for code understanding and completion~\cite{wang2023codet5+}, which is closely related to our domain), as it obtained the best performance.
Also, compared to other encoder-decoder base models, such as the evaluated CodeT5p 6B, it is smaller and thus more affordable in terms of resources and training time. 
However, one shortcoming of CodeT5p 770M is its upper bound of 512 tokens. 
In $DS_{labeled}$, about 83\% of the 276,785 instances are less than 512 tokens. 
These 229,989 KCFs and labels, were used in our experiments. 
The rest of our experiments were conducted using the CodeT5p 770M base model only, and the dataset of 229,989 KCFs which are short enough for this base model.

\subsection{Selection of data sources for fine-tuning}\label{subsec:Selecting_data_sources_for_fine_tuning}

To identify the most effective combination of data sources for fine-tuning, we experimented with three data sources:

\begin{itemize}[leftmargin=*]
    \item \textbf{Labeled KCFs}: As in any classification task, model training requires labeled instances which in our use case are KCFs paired with their misconfigs (Sec.~\ref{subsubsec:Collecting_KCFs_and_labeling_them}).
    \item \textbf{Unlabeled KCFs}: This data source was included as a potential of enhancing the model's ability to understand the typical structure and patterns inherent in KCFs in general (see Sec.~\ref{subsec:Data_preprocessing}).
    \item \textbf{Free text}: This dataset which comprises 232 documents (700MB) obtained by targeted web crawling, includes various K8s-related material, such as the CIS Benchmarks~\cite{cisBenchmark}, authoritative literature, and official documentation~\cite{KubeDoc, kubernetesDoc}.
    This dataset is available in our github repository. 
    It was included as a potential means of broadening \nameFirstComponent's comprehension of K8s-specific language and concepts.
    For that, we explored the use of domain adaptation strategies utilizing CL methods, including techniques such as next sentence prediction (NSP) and masking (see Sec.~\ref{subsec:Large_language_models}). 
\end{itemize}

To identify the optimal data sources and training pipeline that maximize model performance via fine-tuning, we trained four distinct models using the CodeT5p 770M base model. 
In each setup, the labeled KCFs are used in the last fine-tuning step. 
As the results presented in Table~\ref{tab:data_sources_for_fine_tuning} show, first using unlabeled KCFs and then labeled KCFs resulted in the best performance (maximum mean and/or minimum standard deviation of precision, recall, and F1). 
This suggests that preliminary exposure to the general structure of KCFs, followed by pairs of KCFs and their misconfig labels, slightly improves the model's ability to detect misconfigs, compared to fine-tuning using labeled KCFs only. 
Conversely, the introduction of free text appeared to degrade performance slightly, potentially due to its unstructured nature conflicting with the LLM’s intended structured inputs (i.e., KCFs). 
Therefore, it was not included in our proposed method (Sec.~\ref{sec:proposed_method}), which also saves significant amount of time in training (fine-tuning) the model. 
Notably, adding unlabeled KCFs to free text resulted in improved performance compared to fine-tuning without unlabeled KCFs. 
Still, it is evident that the best performance (with minimal input overhead) is achieved when fine-tuning is performed using only KCFs, both unlabeled and labeled.

\begin{table}[!t]
    \resizebox{\columnwidth}{!}{
    \centering
    \begin{tabular}{llll}
        \toprule
        Data sources & Precision & Recall & F1\\
        \midrule
        Labeled KCFs & $0.983 \pm $0.026	 & $0.995 \pm $0.043	 & $0.989 \pm $0.039\\
        Unlabeled KCFs $\rightarrow$ labeled KCFs & $\textbf{0.984} \pm $\textbf{0.027} 
        & $\textbf{0.995} \pm $\textbf{0.037}	 
        & $\textbf{0.989} \pm $\textbf{0.035}\\
        Free text $\rightarrow$ labeled KCFs& $0.979 \pm $0.031 & $0.988 \pm $0.059	 & $0.982 \pm $0.058 \\
        Free text $\rightarrow$ unlabeled KCFs $\rightarrow$ labeled KCFs & $0.983 \pm $0.027	 & $0.995 \pm $0.049 & $0.988 \pm $0.048\\
        \bottomrule
    \end{tabular}
    }
    \caption{Performance of \nameFirstComponent models fine-tuned using various combinations of data sources}
    \label{tab:data_sources_for_fine_tuning}
\end{table}

\subsection{Comparison to existing RB tools}\label{subsec:comparing_with_existing_rule_based_tools}

Having selected the encoder-decoder as our architecture (Sec.~\ref{subsec:Selecting_the_modeling_architecture_and_base_model}), CodeT5p 770M as our base model (Sec.~\ref{subsec:Selecting_the_modeling_architecture_and_base_model}), and the combination of unlabeled and labeled KCFs as our data sources (Sec.~\ref{subsec:Selecting_data_sources_for_fine_tuning}), we conducted experiments for hyperparameter tuning, such as the learning rate and the weight decay. 
We restricted the LoRA adaption~\cite{hu2021lora} to 12.5\%, r=128, and LoRA\_alpha=256. 
The complete list of the tuned hyperparameters used to assess the performance of our method is provided in our GitHub repository.

Using the optimized hyperparameters and the same training set (comprised of unlabeled KCFs followed by labeled KCFs), we retrained \nameFirstComponent's LLM and evalauted it on the test set.
The results of our evaluation of its performance, along with that of industry-standard RB tools for KCF misconfig detection (KubeLinter, Checkov and Terrascan), are presented in Table~\ref{tab:existing_rule_based_tools_comparing}.  
To ensure a fair comparison, each RB tool was evaluated for its ability to detect the \emph{specific subset of misconfigs it is designed to identify}. 
This approach highlights each tool's strengths without penalizing the tools for undetected misconfigs that fall outside their configured rules.
We also evaluate the performance of an ensemble of the three tools (referred to as RB-Ensemble), which covers all of the misconfig subsets they address; this serves as a benchmark for the maximum detection capabilities available.

\begin{table}[ht]
    \resizebox{\columnwidth}{!}{
    \centering
    \begin{tabular}{lllll}
        \toprule
        Tool & Precision & Recall & F1 & \#Labels\\
        \midrule
        \nameFirstComponent& 0.990$\pm$0.020 & 0.999$\pm$0.026 & 0.994$\pm$0.027 & 169 \\
        KubeLinter & $1.0 \pm $0.0 & $0.827 \pm $0.355	 & $0.837 \pm $0.351 & 55\\
        Checkov & $1.0 \pm $0.0 & $0.851 \pm $0.316 & $0.866 \pm $0.319 & 121\\
        Terrascan & $1.0 \pm $0.0 & $0.587 \pm $0.471	 & $0.599 \pm $0.474 & 41\\
         \midrule
         RB-Ensemble & $1.0 \pm $0.0  & $1.0 \pm $0.0 & $1.0 \pm $0.0 & 169 \\
        \bottomrule
    \end{tabular}
    }
    \caption{Performance of various tools for KCF misconfig detection: \nameFirstComponent vs. individual industry-standard RB tools vs. an ensemble of these RB tools}    \label{tab:existing_rule_based_tools_comparing}
\end{table}

As seen in Table~\ref{tab:existing_rule_based_tools_comparing}, compared to each RB tool, \nameFirstComponent demonstrates superior capabilities on the recall and F1 metrics, with precision that is essentially equivalent
. 
Moreover, \nameFirstComponent’s demonstrated ability to detect a larger amount of KCF misconfigs (169 in total) than any of the evaluated RB tools individually, reflects its comprehensiveness and applicability. 
Unlike RB tools that are confined to a limited number of specific misconfigs and achieve perfect precision but low recall, \nameFirstComponent’s approach enables it to identify a large amount of both common and rare misconfigs across various scenarios, with high precision and high recall.

\subsection{\nameFirstComponent error analysis}\label{subsec:error_analysis}

In order to better understand \nameFirstComponent's FPs (shown in Table~\ref{tab:existing_rule_based_tools_comparing}), we manually inspected them with the help of a K8s security expert. 
FP misconfigs appeared in 563 KCFs in the test set (note that a KCF may contain one or more misconfigs). 
Since we could not manually inspect all of these KCFs, we randomly sampled 100 of them, which contain 489 FPs.
Our K8s expert thoroughly examined these KCFs and found that $408/489=83.44\%$ of the FP misconfigs were actually TPs.
These results indicate that \nameFirstComponent's precision is actually higher than shown in Table~\ref{tab:existing_rule_based_tools_comparing}. 
This also means that \nameFirstComponent succeeded in effectively generalizing its knowledge base, providing correct insights that surpass the conventional ground truth in its field, set by the ensemble of three industry-standard RB tools. 
More specifically, \nameFirstComponent succeeded in detecting misconfigs that were not caught by any of the RB tools, since they were \emph{variations} of detection rules and thus didn't entirely fit any coded rule.

Appendix~\ref{app:Misclassified_FPs_examples} provides examples of two cases of misconfig detections provided by \nameFirstComponent that were initially categorized as FPs and later determined by the K8s expert to actually be TPs. 
For each of these two examples, we present the relevant section of the KCF along with a brief description of the misconfig detected by \nameFirstComponent but missed by the RB ensemble. 
A possible reason for \nameFirstComponent's advantage over the RB tools in detecting such misconfigs may stem from both the inherent complexity of KCFs and the advanced capabilities of LLMs; the latter enables the models to learn complex representations of input data~\cite{wang2020recent}, thereby enhancing the ability to identify subtle misconfigs that other non-LLM-based tools might miss. 
LLMs' capacity to generalize from large datasets~\cite{neyshabur2017exploring} may also contribute to their effectiveness in diverse and complex environments such as KCFs.

\subsection{Training set's size}\label{subsec:Sensitivity_analysis}

To characterize \nameFirstComponent's learning curve with respect to the size of its training set, we incrementally increased the training set's size, ranging from 50 to 2,000 misconfig instances \emph{for each misconfig label}.
In cases in which there was a limited quantity of instances for a misconfig, all available instances were used.
Each time, we fine-tuned \nameFirstComponent from scratch, using the optimized hyperparameters described throughout this section, and evaluated its performance using the same test set used thus far.

\begin{figure}[t]
    \centering
    \includegraphics[width=0.98\linewidth]{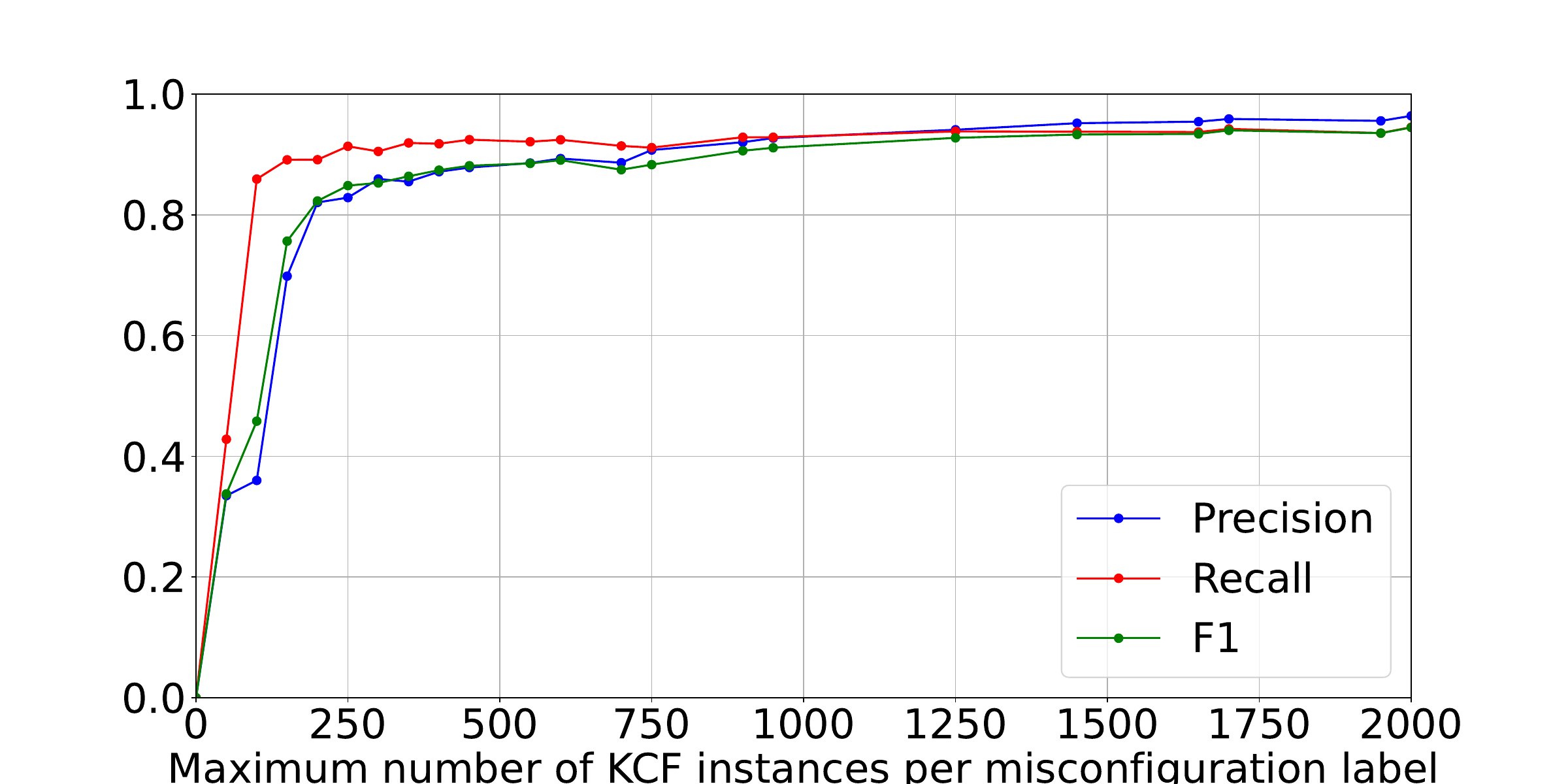}
    \caption{Learning curve analysis for \nameFirstComponent based on incremental increases in the training set's size.}
    \label{fig:sensetivity_analysis}
    
\end{figure}

As can be seen in Fig.~\ref{fig:sensetivity_analysis}, \nameFirstComponent's performance metrics increase rapidly from 0 to around 250 instances and then reach a plateau, converging when there are approximately 1,250 instances from each misconfig label. 
It is important to note that while larger high-quality datasets typically contribute to better model generalization and performance in the realm of DL in general, and LLM specifically~\cite{alom2019state,thambawita2021impact,dong2021survey}, the rate of improvement decreases, eventually becoming marginal.
Thus, while our identified threshold of 250 misconfigs instances serves as an effective minimum, larger datasets may still offer incremental benefits. 

\subsection{Adaptation to new misconfigs}\label{subsec:Adapting_to_new_misconfigs}

The set of known KCF misconfigs is likely to expand occasionally (mainly when a new configuration-related vulnerability is discovered). 
To prevent exploitation of a newly-discovered misconfig, detection tools must be adaptable. 
Traditionally, such adaptation has relied on programmers' in-depth understanding of the detection tool's source code, K8s proficiency, and ability to develop new detection rules or methods to identify the new misconfigs. 
This process of adaptation to newly-discovered misconfigs is time-consuming and burdensome, which limits the responsiveness of RB tools. 
In contrast, with an LLM-based detection method like ours, all that is needed is further fine-tuning of the trained LLM using a small set of KCFs labeled with the newly-discovered misconfig.

\begin{algorithm}[ht] 
\caption{Empirical assessment of \name's adaptability.}\label{alg:adaptability}
  \footnotesize
\begin{algorithmic}[1]
\Require $DS^{adaptation}_{fixed}, LLM^{trained}, DS^{test}$
\State $M \gets (51, 97, 103, 139, 140)$
\State $S \gets (1, 2, 5)$
\ForEach {$m \in \mathcal M $}
\ForEach {$s \in \mathcal S $}
\ForEach {$i \in \mathcal (1, 2, \dots, 10) $}
\State $DS^{adaptation}_{m, s, i} \gets DS^{adaptation}_{fixed} \cup KCF\_sample(m, s, i)$
\State $LLM_{m, s, i} \gets fine\_tune(LLM^{trained}, DS^{adaptation}_{m, s, i})$
\State $precision_{m, s, i}, recall_{m, s, i} \gets evaluate(LLM_{m, s, i}, DS^{test})$
\EndFor
\\
\hskip3em \Return {$Mean(precision_{m, s, i}), St.Dev.(precision_{m, s, i}$)}\\
\hskip3em \Return{$Mean(recall_{m, s, i}), St.Dev.(recall_{m, s, i}$)}
\EndFor
\EndFor
\end{algorithmic}
\end{algorithm}

To empirically assess \nameFirstComponent's ability to adapt to new misconfigs, we devised an experiment (outlined in Alg.~\ref{alg:adaptability}) in which we intentionally omitted a set of five randomly selected misconfig\_IDs, denoted as $M$, from the training set. 
These omitted misconfig\_IDs simulate newly-discovered misconfigs, unseen during training. \nameFirstComponent's LLM (denoted as $LLM^{trained}$), which was trained without any $M$-labeled KCFs, simulates an operational detector that is required to adapt as quickly and accurately as possible. 
We used the same test set as before (denoted as $DS^{test}$), however to expedite the multiple sub-experiments described next we only used 30\% of the data to train and fine-tune $LLM^{trained}$, as described in Sec.~\ref{subsec:Selecting_the_modeling_architecture_and_base_model}. 
For each misconfig\_ID $m \in M$ and sample size $s \in (1, 2, 5)$, in each iteration $i \in (1, 2, \dots, 10)$ we took a random sample of $s$ KCFs labeled with the `new' misconfig\_ID $m$.
This sample, denoted as $KCF\_sample(m, s, i)$, was then merged with a randomly sampled fixed set of 50 `old' KCFs (denoted as $DS^{adaptation}_{fixed}$), resulting in an adaptation dataset $DS^{adaptation}_{m, s, i}$ comprised of labeled KCFs of both old and new misconfig labels. 
This approach was chosen to prevent the fine-tuned model in each iteration $LLM_{m, s, i}$ from being biased towards the new misconfig\_ID $m$. 
After each iteration, defined by the unique combination of $m$, $s$, and $i$, the precision and recall of $LLM_{m, s, i}$ were evaluated using $DS^{test}$. 
Then, for each combination of $m$ and $s$, the mean and standard deviation of the precision and recall were calculated. 
The results of this experiment are presented in Table~\ref{tab:precision_recall_results}.

\begin{table}[h]
    \centering
    \resizebox{\columnwidth}{!}{
        \begin{tabular}{c|c|c|c|c|c|c}
        \hline
         & & \multicolumn{3}{c|}{\textbf{Fine-tuning (for adaptation)}} & \multicolumn{2}{c}{\textbf{Full training}} \\
        \hline
        \textbf{} & $\bm{m}$ & \textbf{$\bm{s}$=1} & \textbf{$\bm{s}$=2} & \textbf{$\bm{s}$=5} & \textbf{30\% data} & \textbf{80\% data} \\
        \hline
        \multirow{5}{*}{\rotatebox{90}{\textbf{Precision}}} & 51 & 0.083±0.180 & 0.444±0.416 & 0.445±0.113 & 1 & 1\\
        & 97 & 0.433±0.498 & 0.263±0.404 & 0.366±0.090 & 0.579 & 0.923\\
        & 103 & 0.475±0.506 & 0.817±0.317 & 0.857±0.000 & 1 & 1\\
        & 139 & 0.875±0.317 & 0.944±0.118 & 0.943±0.074 & 1 & 1\\
        & 140 & 0.994±0.013 & 0.951±0.046 & 0.942±0.040 & 0.967 & 0.972\\
        \hhline{=======}
        \multirow{5}{*}{\rotatebox{90}{\textbf{Recall}}} & 51 & 0.005±0.010 & 0.048±0.054 & 0.256±0.153 & 1 & 1\\
        & 97 & 0.042±0.044 & 0.083±0.096 & 0.467±0.143 & 0.917 & 1 \\
        & 103 & 0.167±0.208 & 0.574±0.374 & 1 & 1 & 1\\
        & 139 & 0.417±0.180 & 0.685±0.242 & 1 & 1 & 1\\
        & 140 & 0.826±0.103 & 0.956±0.045 & 0.969±0.076 & 1 & 1\\
        \hline
        \end{tabular}
    }
    \caption{\nameFirstComponent's precision and recall for selected misconfig\_IDs ($\bm{m}$) as a function of the size of the sample ($\bm{s}$) in fine-tuning (left side) or the size of the training set in full training (right side).}
    \label{tab:precision_recall_results}
\end{table}

Table~\ref{tab:precision_recall_results} shows a promising learning curve in the context of \nameFirstComponent's ability to quickly and accurately adapt to newly-discovered KCF misconfigs. 
The empirical results show that to achieve a precision level of 0.8 or more, only two {m}-labeled KCFs (i.e., {s}=2) are required in 60\% of cases and just one {m}-labeled KCF is needed in 40\% of cases. 
Satisfactory recall levels require slightly more {m}-labeled KCFs, however in 60\% of cases, five {m}-labeled KCFs are sufficient to achieve near-zero FN rates. 
These empirical results demonstrate our LLM-based method's adaptability and effectiveness in continuously evolving environments, which are maintained in such settings \emph{without compromising \nameFirstComponent's performance} or increasing the demands on RB tool developers.

When looking closer at the results presented in Table~\ref{tab:precision_recall_results} we can see that some misconfig\_IDs (i.e., 51 and 97) suffer from many FNs and FPs when only a handful of them are used to further fine-tune a trained LLM. 
In further analysis of these misconfig\_IDs, we found that both of the associated misconfigs are extremely similar to other misconfigs, and thus they tend to be misclassified. 
More specifically, misconfig\_ID 97 is defined under `securityContext/capabilities/\emph{add},' while several other misconfig\_IDs are defined under `securityContext/capabilities/\emph{drop}.'
This makes it difficult for the LLM, which has already learned from the old misconfigs, to distinguish between them based such a small sample of new KCFs. 
Similarly, the fine-tuned LLM frequently failed to recognize misconfig\_ID 51 and confused it with another misconfig\_ID, because they are both associated with role-based access control (RBAC). 
Still, as can be seen 
on the right side of Table~\ref{tab:precision_recall_results}, when \nameFirstComponent is trained with more data (30 or 80\% of the $M$-labeled KCFs), the precision and recall using $DS^{test}$ converge to one.
Also, the gap between having a small or large set of $m$-labeled KCFs (denoted as $KCF\_{sample(m)}$) can be bridged by initially fine-tuning $LLM^{trained}$ using whatever size of $KCF\_{sample(m)}$ is available. 
Then, using $LLM_{m}$ (i.e., $LLM^{trained}$ which was further fine-tuned using $DS^{adaptation}_{m, s}$), we can scan other KCFs, either locally available or obtained on the Internet to gather additional $m$-labeled KCFs. 
With this approach, even an $LLM_{m}$ with relatively low recall and/or precision can increase the size of $KCF\_{sample(m)}$ by gathering KCFs that have the newly-discovered $m$ but have not yet been labeled as such. 
Then, a second round of fine-tuning (using the increased $KCF\_{sample(m)}$) would enable the \nameFirstComponent's quick and accurate adaptation to the newly-discovered misconfig.

\subsection{Localization, reasoning, and remediation}\label{subsec:Localizing_detected_misconfig}

Manually validating \nameSecondComponent's output for each misconfig detected is a time-consuming task that should be performed by an expert.
Therefore, a K8s security expert validated \nameSecondComponent's performance on a random sample of 30 detected KCF misconfigs (this took approximately six minutes for each misconfig). 
The K8s expert confirmed that \nameSecondComponent excels in these explanatory tasks, achieving a score of 30/30 in terms of both its explanations and remediation suggestions. 
In terms of localization, in 20 of the 30 cases, \nameSecondComponent correctly identified that necessary lines were missing in the provided KCFs, and therefore could not provide a line number. 
In the remaining 10 cases, \nameSecondComponent accurately pinpointed the issues in the correct lines.

\section{Discussion}\label{sec:discussion}

\subsection{Lessons learned}\label{subsec:Lessons_learned}

During our research, the following valuable insights emerged. 
First, similar to  previous research~\cite{lanciano2023analyzing, minna2024analyzing}, using the basic prompt engineering and few-shot learning techniques to adapt pretrained LLMs for KCF misconfig detection results in relatively poor performance (Sec.~\ref{subsec:Evaluating_Existig_pre_trained_LLMs}). 
For this highly specific task of multi-label classification within (semi-structured) KCFs, we found that \emph{a thoroughly optimized and fine-tuned LLM is capable of achieving near-perfect performance}, with precision equivalent to industry standards and superior recall. 
In comparison, for the task of KCF misconfig localization, reasoning, and remediation, our experimental results (validated by a K8s expert) showed that few-shot learning with a pretrained model is sufficient. Hence, it seems that \emph{reconstructing specific misconfig\_IDs is much more difficult for an LLM than generating less restrictive text for misconfig localization, reasoning and remediation}. 

Second, it seems that with a suitable LLM architecture, a domain-relevant base model~\cite{wang2023codet5+}, and optimized hyperparameters, \emph{relatively-small quantities of training data are sufficient for \nameFirstComponent to achieve satisfactory performance}. 
This applies both to the initial fine-tuning whose learning curve achieves high precision and recall (about 83\% and 91\%, respectively) when there are 250 KCFs per misconfig (Fig.~\ref{fig:sensetivity_analysis}), and to the adaptation process, which, in most cases, requires only a handful of relevant KCFs in order to be able to detect a newly-discovered misconfig (Table~\ref{tab:precision_recall_results}).

A third lesson learned during our experimentation is that \emph{when an LLM is used for classification, encoding (compressing) the textual label is beneficial}. 
Most importantly, with a given amount of labeled KCFs in the training set, predicting an encoded misconfig label may improve the classification performance (longer texts are harder to reconstruct). 
Furthermore, a reduction in the number of processed tokens (during \nameFirstComponent's training and inference) directly translates to accelerated and less expensive compute.

A fourth insight, which was counter-intuitive for us, is that \emph{during \nameFirstComponent's fine tuning, focusing exclusively on KCFs maximizes the precision, recall, and F1} (Table~\ref{tab:data_sources_for_fine_tuning}). 
That is, incorporating free-text information about K8s along with KCFs does not enhance performance on these metrics, but rather actually impairs it slightly. 
This finding suggests that (1) when the LLM is designed to analyze structured data (i.e., KCFs), there is no use in training it using unstructured data; (2) fewer inputs (i.e., KCFs only) are sufficient for achieving accurate outputs; and (3) labeled KCFs are essentially sufficient for realizing the LLM's potential in detecting KCF misconfigs, as the contribution of unlabeled KCFs is negligible.


\subsection{Research limitations}\label{subsec:Research_limitations}

In our empirical evaluation, the best-performing base LLM (CodeT5p 770M) places a 512 token limit, such that only 83\% of the KCFs in our dataset are covered. 
Nevertheless, CodeT5p 770M may be replaced by a less restrictive base LLM in the future. 
Another limitation is that during error analysis (see Sec.~\ref{subsec:error_analysis} and Sec.~\ref{subsec:Localizing_detected_misconfig}), the expert manually only analyzed a sample of 100 KCFs which contained alleged FPs, and 30 randomly selected outputs of \nameSecondComponent. 
While not all of the KCFs were examined in our manual error analysis, a non-negligable portion of those examined demonstrated \nameFirstComponent's ability to outperform existing RB tools and \nameSecondComponent's ability to produce valuable and correct responses.

\section{Related work}\label{sec:related_work}

\subsection{The use of LLMs in static analysis}\label{subsec:LLMs_in_static_analysis}

LLMs such as ChatGPT~\cite{achiam2023gpt}, Bard~\cite{Bard}, Gemini Pro~\cite{pichai2023introducing}, GPT-4, and GPT-3.5 have been leveraged extensively in the software and hardware security analysis domain, uncovering both promising applications and notable limitations. 
Several recent studies~\cite{kwon2023exploring,yu2024security,ahmad2024hardware,liu2023harnessing,li2023assisting} demonstrated LLMs' efficacy in static analysis tasks like correcting defective Ansible scripts, detecting code security defects, detecting and repairing security bugs in hardware description languages, and automating static binary taint analysis. 
These tasks are similar to our task of KCF misconfig detection in that they use LLMs for static analysis. 
These studies theoretically suggested the need to fine-tune LLMs for specific applications to optimize performance. 
However in contrast to those studies, we evaluated the impact of fine-tuning on performance quantitatively.
Other studies~\cite{petrovic2023machine,lian2023configuration} examined the ML's integration with runtime DevSecOps and the automation of configuration validation. 
These studies not only underscored the adaptability of generative AI in security contexts but also emphasized the need for further refinement, such as fine-tuning, to fully exploit generative AI technologies' adaptability in these complex applications. 
Our research contributes uniquely to this field; to the best of our knowledge, we are the first to employ a fine-tuned LLM specifically for robust and cost-efficient static analysis of KCFs, detecting KCF misconfigs, and providing detailed reasoning regarding the misconfigs, their exact localization, and remediation suggestions.

\subsection{KCF misconfig detection}\label{subsec:KCF_misconfig_detection} 

Recently, topology graphs were proposed~\cite{blaise2022stay} as a means of revealing attack vectors in security configurations and assigning them security risk scores. 
This approach helps detect KCF misconfigs by mapping the K8s' deployment, identifying vulnerabilities, and highlighting security weaknesses. 
However, it faces challenges in scalability and practical application, requires in-depth K8s expertise, occasionally fails to scan certain containers, and lacks automated remediation suggestions. 
In another study~\cite{rahman2023security}, SLI-KUBE, an innovative RB tool for KCF misconfig detection, was proposed. 
However, in terms of coverage, SLI-KUBE can only detect 11 specific misconfigs using its predefined rule set, while our method was empirically shown capable of detecting 169 KCF misconfigs (see Sec.~\ref{sec:Experimental_results}). 
In addition, our method's use of fine-tuning enables it to quickly and easily adapt to newly-discovered misconfigs, based on just a few KCFs labeled with the new misconfig (Sec.~\ref{subsec:Adapting_to_new_misconfigs}). 
Fine-tuning also offers a substantial advantage over traditional RB tools
, which suffer from a non-negligible incidence of FPs~\cite{minna2024analyzing} and are limited due to their reliance on static rule sets which do not adapt to new misconfigs without manual updates.

\subsection{KCF misconfig remediation}\label{subsec:KCF_misconfig_remediation}

While existing tools can detect KCF misconfigs, they often lack the ability to suggest effective corrective measures. 
Our method is unique, since in addition to detecting misconfigs in KCFs, it also provides detailed reasoning, as well as the specific locations of detected misconfigs and remediation. 
This aids greatly in understanding the root cause of misconfigs, enabling more effective and targeted fixes.

In a preliminary study~\cite{minna2024analyzing}, LLMs' potential in correcting security misconfigs in K8s Helm charts was explored. 
Existing RB tools (similar to the ones used in our study) were employed to detect misconfigs; then LLMs were used to suggest mitigation steps aimed at enhancing configuration. 
Thus, unlike in our study, LLMs were not considered a direct replacement for RB tools but rather as a means of correcting any misconfigs they detect (potentially inadvertently introducing new misconfigs in their attempt to correct others). 
Moreover, the LLMs used in that study are ChatGPT and Gemini, which are general-purpose pretrained LLMs that have not undergone fine-tuning.
Another drawback is that sending files for inspection by external LLMs via API calls, as done in that study, poses security risks~\cite{diaz2021web,ariffin2020api} and adds unnecessary costs. 
To overcome these issues, in our research, we used an LLM developed by Mistral~\cite{jiang2023mistral} which is open-source (and thus free of charge), does not require external API interactions, and can be adapted to meet our specific needs. 
In addition, in the evaluation of this preliminary study, the authors used a limited sample of only 60 Helm charts, whereas we collected and labeled over 276,000 KCFs to evaluate our proposed method. 

Another study~\cite{lanciano2023analyzing} explored LLMs' potential for analyzing KCFs. The authors proposed a pipeline to identify whether the KCF is of `good' or `bad' quality. 
They used Polaris~\cite{Polaris} to identify one specific misconfig
, and leveraged an LLM to perform few-shot learning, assessing the presence of this misconfig in other KCFs, and recommending best practices for developers. 
Despite similarities to our research, some limitations and differences between the two studies are worth noting: (1) The authors used a small dataset of only 100 KCFs, which is much smaller than our dataset of over 276,000 KCFs. 
(2) For training, the authors relied on few-shot learning~\cite{brown2020language,lin2022few,zhang2023machine}, which, as they acknowledged, restricts the LLM's ability to learn and generalize complex relationships from the data. 
In contrast, we performed fine-tuning with a significantly larger dataset, allowing the model to better capture and generalize complex relationships.
(3) They focused on binary classification~\cite{kolo2011binary}, determining the presence or absence of a specific misconfig in KCFs, thus providing limited feedback to developers; in contrast, our method is trained to detect up to 169 misconfigs.
(4) Our approach not only identifies misconfigs but also pinpoints their location, explains their reasons, and offers targeted suggestions for remediation.
(5) Unlike the prior study, which did not include results and evaluation metrics, we have comprehensively described our experimentation, evaluation, and findings, and publicly share our data and code.

In another study whose goals are also relatively similar to ours~\cite{haque2022kgsecconfig}, the authors proposed a completely different approach. 
Specifically, they suggested the use of free text documents (mostly from blog posts) to infer a knowledge graph, which was used to identify misconfigs and provide remediation. 
However, unlike our research, their focus was on matching a given textual vulnerability description to a configuration concept, i.e., extracting knowledge from free text.
A shortcoming in this research noted by the authors is that the method's performance relies heavily on the free text documents' comprehensiveness, as well as on the quality of their text processing procedure; our proposed method does not face these limitations. 
In addition, their evaluation was performed on a much smaller scale than ours, and configuration concepts were manually labeled by the researchers, a method which the authors acknowledged as subjective (unlike the vast amount of KCFs we labeled using three unbiased industry-standard RB tools).

\section{Conclusion}\label{sec:conclusion}

In this research, we explored LLMs' ability to enhance the security of K8s systems. 
For that, \name does more than merely detect (a wide variety of) KCF misconfigs; it provides their exact location in the inspected KCFs, provides clear reasoning from which the KCF developer can learn, and offers actionable recommendations for remediation.
\name does not rely on external APIs, thus ensuring that KCFs are not exposed on the web, enhancing security; this approach also reduces costs. 
Unlike previous studies, we developed a comprehensive LLM-based method, and quantitatively evaluated it.
Our extensive evaluation demonstrated \name's excellent performance, validated by a K8s expert. 
In addition to contributing to containers' security via mitigating KCF misconfig-related risks, by making our data (including the UMI) and code freely accessible, this work will contribute to future advancements in the field, specifically in terms of standardization, benchmarking, and security-oriented, LLM-based analysis and remediation methods.

Future work in this domain can focus on (1) automatically generating a misconfig-free version of the input KCF to eliminate the detected misconfigs, and (2) enhancing \name's ability to prioritize the detected misconfigs based on their severity level.

\clearpage

\bibliographystyle{ACM-Reference-Format}
\bibliography{simple_kube_bib}

\appendix

\section{Entity matching prompt}\label{app:Entity_matching_prompt_structure}

\begin{figure}[H]
\centering
    \includegraphics[width=1\linewidth]{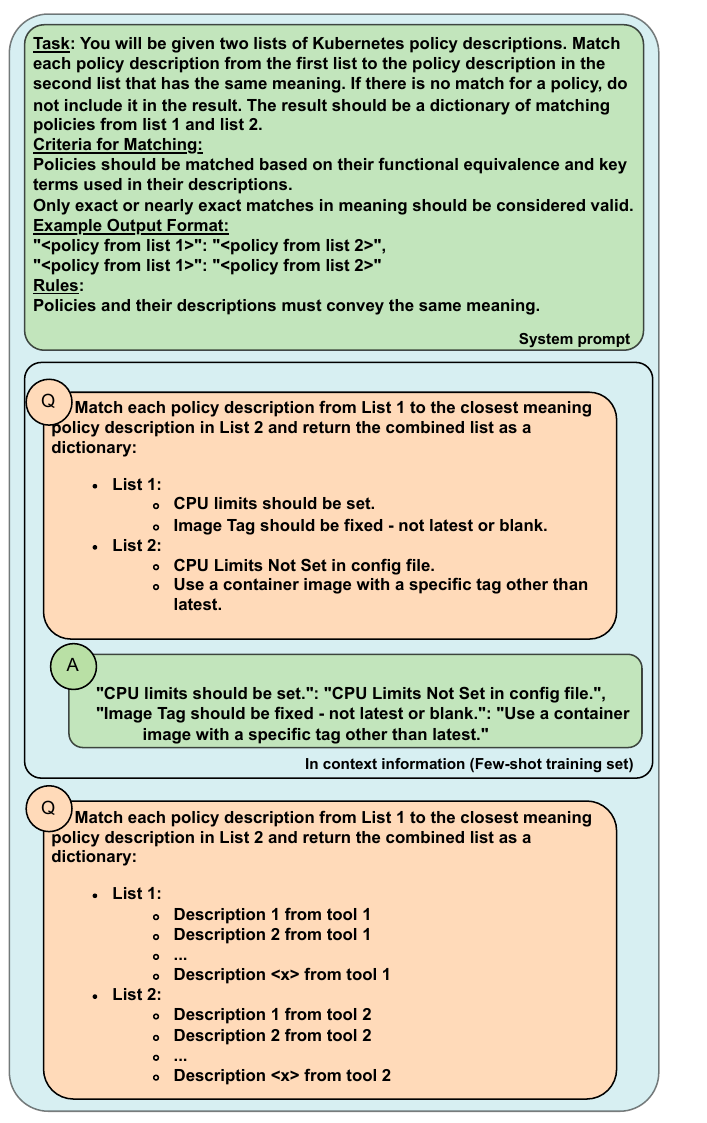}
\caption{Entity matching prompt, used during the UMI creation process, as part of \nameFirstComponent's preprocessing component.}
\label{figbox:entity_matching_prompt}
\Description{This figure presents a task prompt for matching Kubernetes policy descriptions from two lists based on functional equivalence and key terms. The task involves:

Instructions: Match each policy description from the first list to the same meaning policy description in the second list. Exclude unmatched policies.
Criteria for Matching: Policies must convey the same meaning and be nearly exact in terms of functional equivalence and key terms.
Example Output Format: The result should be a dictionary of matched policies from both lists, e.g., { "CPU limits should be set.": "CPU Limits Not Set in config file." }.
Example:

List 1:
CPU limits should be set.
Image tag should be fixed - not latest or blank.
List 2:
CPU limits not set in config file.
Use a container image with a specific tag other than latest.
The task aims to match policy descriptions accurately and return the result as a dictionary.}
\end{figure}

Figure.~\ref{figbox:entity_matching_prompt} outlines the structure of the optimized prompt we used for entity matching while creating the UMI (Sec.~\ref{subsubsec:creating_a_unified_misconfig_index}. This prompt has three segments: (1) a system prompt which describes the entity matching task, the criteria for matching, and the structure of the required output; (2) a few-shot training set; and (3) new lists comprising entities to be matched. We used this prompt in order to map all of the variations
of a misconfig (from three different RB tools) to a single ID.

\section{Encoded and decoded class labels}\label{app:Example_of_Encoded_and_decoded_labels}

\begin{figure}[H]
\centering
\includegraphics[width=1\linewidth]{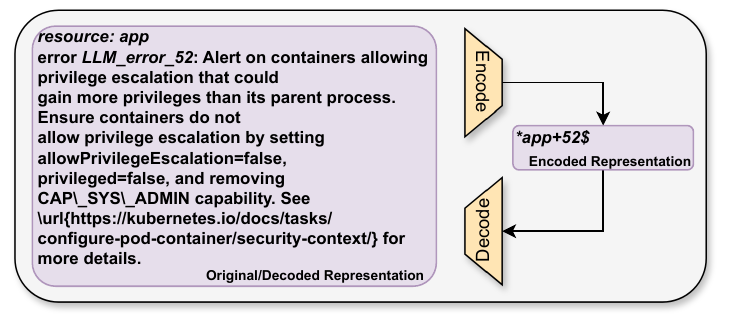}
\caption{Encoded and decoded representations of an example misconfig class label.}
\label{fig:encoding_example}
\Description{This figure illustrates the encoding process for Kubernetes configuration file (KCF) misconfig descriptions. The table presents two columns: the left column shows the original, decoded misconfig description, while the right column shows its corresponding encoded form.

Decoded/Original Misconfig Description: This column provides a detailed description of the misconfig. For example, it specifies an alert for containers allowing privilege escalation. The description includes specific instructions to prevent privilege escalation by setting allowPrivilegeEscalation=false, privileged=false, and removing the CAP_SYS_ADMIN capability. A reference URL to the Kubernetes documentation is provided for further details.

Encoded: This column shows the encoded representation of the misconfig. In this example, the encoded form is *app+52\$, which succinctly represents the detailed misconfig description from the left column.

This encoding process helps in standardizing and simplifying the representation of misconfig descriptions for efficient processing and analysis.}
\end{figure}

Figure.~\ref{fig:encoding_example} illustrates the encoded and decoded representations of an example misconfig class label (as discussed in Sec.~\ref{subsubsec:Collecting_KCFs_and_labeling_them}). The original/decoded representation contains detailed information about a misconfig related to privilege escalation in K8s containers. This relatively long textual description is encoded into the compressed format: <impacted\_K8s\_resource>+<misconfig\_ID>, in this case app+52. 
Encoding the labels decreases the overall size of labeled KCFs, such that
(1) higher coverage is provided by LLMs that impose size limits, (2)
training is accelerated, and (3) fewer misclassifications are likely to
occur, as shorter class labels are easier for the LLM to (re-)generate. Decoding the labels provides more context to \nameSecondComponent during inference (see Sec.~\ref{subsubsec:decoding_splitting}).

\section{\nameSecondComponent prompt}\label{app:nameSecondComponent_Model_Inputs_and_Outputs_for_Misconfig_Remediation_Localization_and_Explanation}

\begin{figure}[H]
    \centering
    \includegraphics[width=1\linewidth, height=0.8577\textheight]{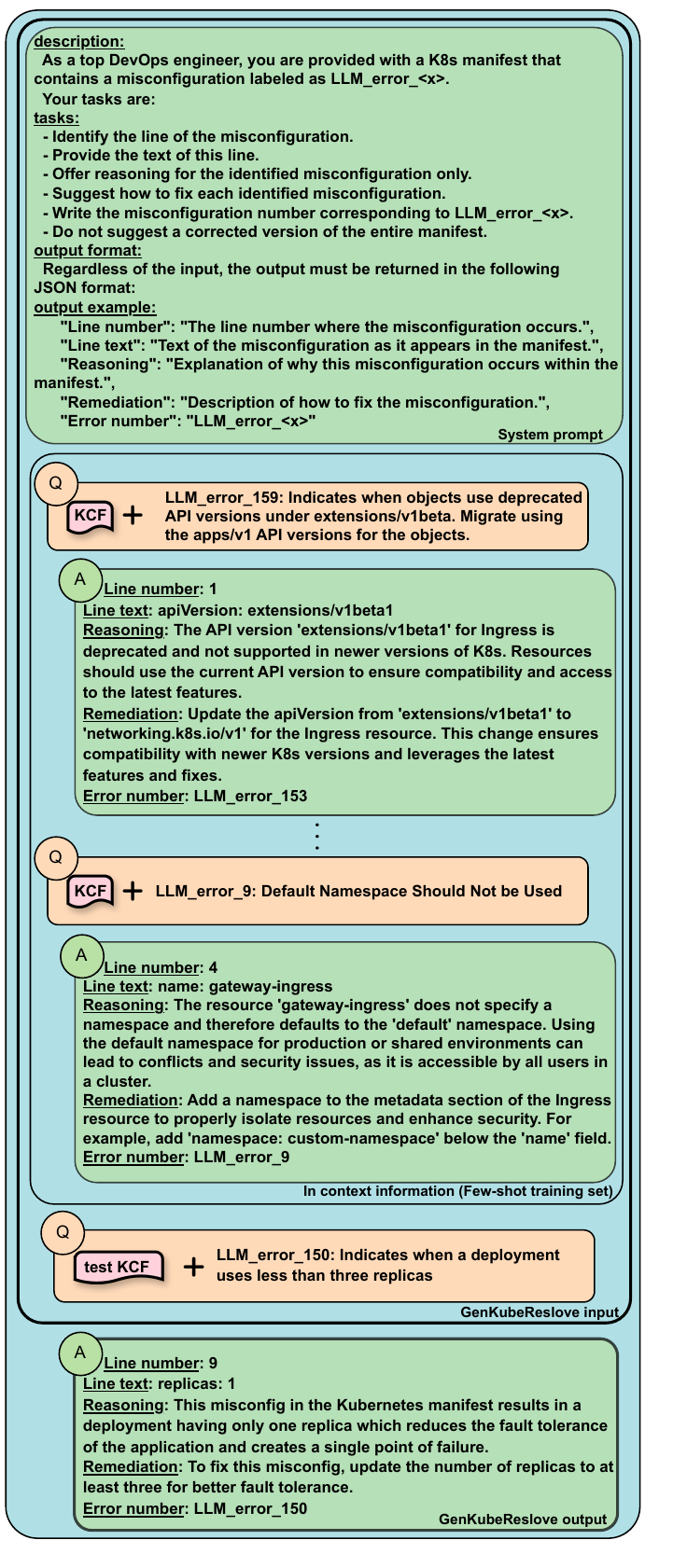}
\caption{The system prompt devised for the Mistral LLM and an example of a few-shot learning process for localization, reasoning, and remediation of a detected KCF misconfig.}\label{figbox:reasoning}
\Description{This figure presents a system prompt for the Mistral LLM and an example of a few-shot learning process to localize, reason, and remediate detected KCF misconfigs.

Prompt Description:

Tasks for the model include identifying the line of the misconfiguration, providing the text of this line, offering reasoning, suggesting fixes, and noting the error number.
The output must be in JSON format, detailing line number, line text, misconfig explanation, fix suggestion, and error number.
<Few-Shot Learning Examples>
This prompt and example demonstrate how Mistral LLM assists in accurately identifying, explaining, and fixing KCF misconfigs.}
\end{figure}

Figure.~\ref{figbox:reasoning} outlines the structure of the optimized prompt we used in \nameSecondComponent for providing localization, reasoning and remediation suggestions to each misconfig detected by \nameFirstComponent (see further details in Sec.~\ref{subsubsec:Specializing_a_LLM_for_Localization_reasoning_and_remediation}). This prompt has three segments: (1) a system prompt which describes the required outputs and their format; (2) a few-shot training set; and (3) a test KCF to be processed by \nameSecondComponent. At the bottom we can see \nameSecondComponent's output for this test KCF: (1) the exact location of the detected misconfig is line 9 (the current text within this line is 'replicas:1'); (2) the reason for alerting is that the current configuration results in a single point of failure; (3) for remediation, the KCF developer is advised to increase the number of replicas. 

\section{Prompt used with the general purpose pretrained LLMs for misconfig detection}\label{app:Few_shot_learning_example}

\begin{figure}[H]
\centering
    \includegraphics[width=1\linewidth]{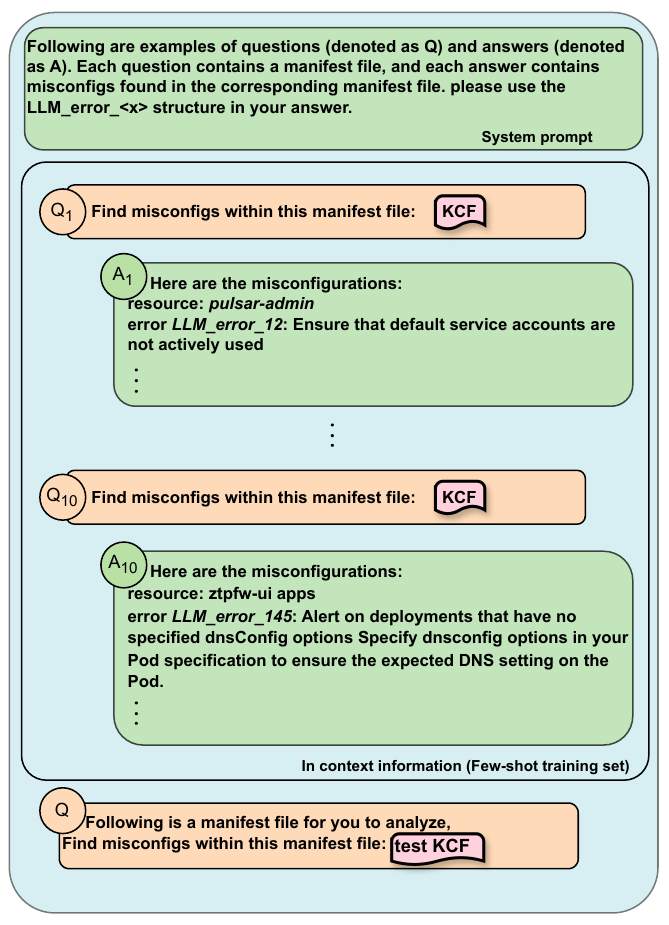}
\caption{General instruction (top), 2 out of the 10 few-shot learning pairs of KCFs and their misconfig labels (middle), and a test KCF for which the LLM is requested to generate misconfig labels (bottom).}
\label{figbox:few_shot_learning}
\Description{This figure presents a few-shot learning example of identifying misconfigs in Kubernetes configuration files (KCFs). The example is divided into three sections:

General Instruction (Top): This section provides an overview of the task, explaining that each question (Q) contains a manifest file, and each answer (A) contains misconfigs found in the corresponding manifest file.

Training Pairs (Middle): This section presents two out of 10 training pairs of KCFs and their corresponding misconfiguration labels. Each pair includes a question (Q) with a manifest file snippet and an answer (A) listing the  misconfigs identified. 

Test KCF (Bottom): This section presents a manifest file for the LLM to analyze and generate misconfiguration labels. The manifest file describes a StatefulSet with metadata and specific match labels for nginx.}
\end{figure}

Figure.~\ref{figbox:few_shot_learning} outlines the structure of the optimized prompt we used with pretrained general-purpose LLMs for them to be able to detect misconfigs in KCFs. This prompt has three segments: (1) a system prompt which describes the task; (2) a few-shot training set; and (3) a test KCF to be processed by the pretrained LLMs. The experimental results eventually showed that more sophisticated LLM training techniques (mostly fine-tuning) are required for reaching satisfactory detection performance (see further details in Sec.~\ref{subsec:Evaluating_Existig_pre_trained_LLMs}).

\section{Misclassified FPs examples}\label{app:Misclassified_FPs_examples}

\begin{figure}[H]
\centering
    \includegraphics[width=1\linewidth]{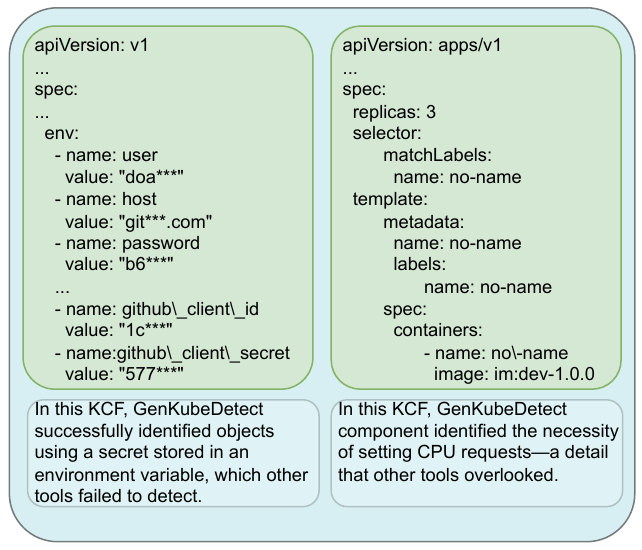}
\caption{Examples of test KCFs that were misclassified as FPs by \nameFirstComponent and subsequently verified by a K8s expert as TPs, thus demonstrating \nameFirstComponent's effectiveness in detecting actual KCF misconfigs.}\label{figbox:new_gt}
\Description{This figure presents two KCFs that were initially misclassified as FPs but later verified as TPs, demonstrating \nameFirstComponent's effectiveness in detecting actual misconfigs.

KCF 1: The manifest includes environment variables with sensitive information. \nameFirstComponent successfully identified objects using a secret stored in an environment variable, which other tools failed to detect.

KCF 2: The manifest specifies a deployment with multiple replicas. \nameFirstComponent identified the necessity of setting CPU requests, a detail overlooked by other tools.

These examples highlight \name's superior capability in accurately detecting and verifying KCF misconfigs.}
\end{figure}

Figure.~\ref{figbox:new_gt} provides examples of test KCFs that which were initially categorized as FPs but were subsequently verified by a K8s security expert to be acutally TPs. 

The first KCF example (left) involves the use of a secret stored in an environment variable. \nameFirstComponent successfully identified the risky usage of secrets, which the RB tools failed to detect. The snippet shows environment variables containing sensitive information like user credentials and GitHub client secrets.

The second KCF example (right) addresses the necessity of setting CPU requests for containers. \nameFirstComponent identified that such setting were not provided, a detail that the RB tools overlooked. The snippet includes a container specification where CPU requests should be defined to ensure proper resource allocation and performance (see Sec.~\ref{subsec:error_analysis}).

\setlength{\fboxsep}{1pt}
\setlength{\fboxrule}{1pt}

\end{document}